\documentclass[
p
reprint,
pre,
showpacs,
showkeys,
longbibliography
]{revtex4-1}
\usepackage[displaymath, mathlines]{lineno}
\usepackage{amsmath, amssymb, graphicx,amsfonts,bm
,hyperref
}

\newcommand{\D}[0]{\mathrm{d}}
\newcommand{\fig}[2]{Fig.~\ref{#1}#2}

\newcommand{\comment}[1]{}

\begin{document}

\title{Phase Transitions in the Quadratic Contact Process on Complex Networks}%

\author{Chris Varghese}
\email{varghese@phy.duke.edu}
\affiliation{Department of Physics, Duke University, Durham, North Carolina, USA}

\author{Rick Durrett}
\email{rtd@math.duke.edu}
\affiliation{Department of Mathematics, Duke University, Durham, North Carolina, USA}


\date{\today}
\begin{abstract}
The quadratic contact process (QCP) is a natural extension of the well studied linear contact process where infected (1) individuals infect susceptible (0) 
neighbors at rate $\lambda$ and infected individuals recover ($1 \longrightarrow 0$) at rate 1. In the QCP, a combination of two 1's is required to effect a $0 \longrightarrow 1$ change. We extend the study of the QCP, 
which so far has been limited to lattices, to complex networks.
We define two versions of the QCP -- vertex centered (VQCP) and edge centered (EQCP) 
with birth events $1-0-1 \longrightarrow 1-1-1$ and $1-1-0 \longrightarrow 1-1-1$ respectively, where `$-$' represents an edge. We investigate the effects of network topology by considering the QCP on random regular, Erd\H{o}s-R\'{e}nyi 
and power law random graphs. We perform mean field calculations as well as simulations to find the steady state fraction of occupied vertices as a function of the birth rate. We find that on the random regular 
and Erd\H{o}s-R\'{e}nyi graphs, there is a discontinuous phase transition with a region of bistability, whereas on the heavy tailed power law graph, the transition is continuous. The critical birth rate is found to be positive 
in the former but zero in the latter.
\end{abstract}

\maketitle

\section{Introduction}
Inspired by technological and social networks, the study of complex networks has seen a surge in the past fifteen years \cite{newman03,boccaletti06,caldarelli07,cohen10,newman10}. 
Research has traditionally progressed in two distinct directions -- dynamics \emph{of} networks and dynamics \emph{on} networks. The former is concerned with the formation of a network or change in its structure with time, whereas the 
latter deals with processes (deterministic or stochastic) taking place on a fixed network. Preferential attachment and its many generalizations \cite{albert02,cooper03} are prototypical examples of the first type. 
Examples of the second are epidemics \cite{pastor01,newman02,berger05,volz08}, the voter model for the spread of an opinion \cite{sood05,durrett06,castellano09}, cascades \cite{watts02,gleeson07,montanari10} 
that model spread of a technology, and evolutionary games \cite{ohtsuki06}. The phase transitions \cite{dorogovtsev08,barrat08} associated  with these models have been of particular 
interest. 

In the mathematics community, spatial models are studied under the heading of \emph{interacting particle systems} \cite{liggett85}.
One of the simplest models of those models is the contact process \cite{harris,henkel09,marro05} (equivalent to the SIS model in epidemiology). 
In the linear contact process each site can be in one of two states which we will call 1 and 0. 0's become 1 at a rate proportional to the number of 1 neighbors they have and 1's become 
0 at a constant rate (here and in all following models, unless otherwise specified, the processes occur in continuous time).

A natural extension of the linear process is the quadratic contact process (QCP) where each $0 \rightarrow 1$ event will require two other sites in state 1. We will occasionally refer to 1 as being the ``occupied'' state and 
0 as being ``vacant'', and 
the events $0\rightarrow 1$ and $1 \rightarrow 0$ to be birth and death events respectively. 
At this stage, the model is quite general in that we do not specify where the two 1's that cause the $0 \rightarrow 1$ event must be located with respect to the 0. 
On the 2D lattice, specifying these locations leads to different realizations of the QCP. For example, Toom's North-East-Center model  (originally defined in discrete time) 
allows a 0 at site $x$ to be filled if its neighbors $x+(0,1)$ and $x+(1,0)$ are occupied \cite{toom80}. 
Chen \cite{chen92,chen94} has studied versions of Toom's model in which two or three specified adjacent pairs or all four adjacent pairs are allowed to reproduce. 
Evans, Guo and Liu \cite{guo07,liu07,guo08,guo09,liu09} have studied the QCP as a model for 
adsorption-desorption on a two dimensional 
square lattice. In the version of the model studied by Liu \cite{liu09}, 0 becomes 1 at rate proportion to the number of adjacent pairs of 1 neighbors. He found a discontinuous phase 
transition with a region of bistability, where the 1's die out starting from a small density. He also
found that by introducing spontaneous births at a sufficiently high rate, the
transition becomes continuous. 

The QCP is similar to Schl\"{o}gl's second model \cite{schlogl72} of autocatalysis characterized by chemical reactions $2X \longrightarrow 3X, X \longrightarrow \o$  
where $X$ represents the reactant. Grassberger \cite{grassberger82} studied a version of 
Schl\"{o}gl's second model in which each site has a maximum occupancy of two and doubly occupied sites give birth to a neighboring vacant site. He found that the model shows a continuous phase transition in 2D.

Studies to date on the QCP have been limited to regular lattices in low dimensions. In this paper, we extend the study to complex networks. 
There are two ways to view the QCP on networks:

\begin{itemize}
 \item as a model that replaces the linear birth rate of the contact process that has been extensively studied on networks \cite{chatterjee09,parshani10}, by a quadratic birth rate.
\item as an alternative model for the spread of rumors, fads and technologies such as smart phones in a social network. In sociology the requirement of more than a single 1 for the ``birth'' event is called 
complex contagion \cite{centola07}. Also related are the threshold contact process \cite{chatterjee13} and models for the study of ``cascades'' \cite{gleeson07}. The key difference
here is that the QCP involves a death event that represents the loss of interest in the fad or technology and the rate for birth events is a function of the actual number and not the fraction of occupied neighbors.
\end{itemize}

The questions we are interested in are: 
How does network topology
affect these phase transitions? What model and network features lead to discontinuous versus continuous phase transitions?

The paper is organized as follows. We define the specific QCP that we study in Section \ref{model} and we do mean field calculations in Section \ref{mean}. In Section \ref{rigor} we present a few rigorous results about the QCP. 
Simulation results are presented in Section \ref{simulation}, followed by some concluding remarks in Section \ref{conclusion}.

\section{Model definition} \label{model}
The birth event in the linear contact process can be formulated as each $1-0$ edge converts to a $1-1$ edge at a constant rate 
$\lambda$. Such a definition can be easily extended to the quadratic case by defining the birth event in terms of connected vertex triples. Two 
such definitions are possible: $1-0-1 \longrightarrow 1-1-1$, and $1-1-0 \longrightarrow 1-1-1$.
We call the former version the vertex centered QCP (VQCP) because the central 0 \emph{vertex} is getting filled by its two neighboring 1s, and the latter as the edge centered QCP (EQCP) as it can be viewed as a 
$1-1$ \emph{edge} giving birth on to a neighboring vacant vertex.
 Note that the models can also be defined in terms of how a vacant vertex gets filled i.e., suppose that a 0 vertex has 
$k$ 1 neighbors and $j$ $1-1$ neighbors \footnote{The number of $1-1$ neighbors of a vertex $x$ is $|\{(y,z):x\sim y, y\sim z, z \neq x, \textrm{state of }y = \textrm{state of }z =1 \}|$ }, 
then the 0 vertex will become 1 at rates $\binom{k}{2} \lambda$ and $j\lambda$ in the VQCP and EQCP respectively. Death events $1 \longrightarrow 0$ occur at rate 1 as in the linear
process.

If the death rate is changed to zero, the VQCP reduces to bootstrap
percolation
\cite{baxter10} 
where vertices that are occupied remain occupied forever and vacant vertices that have at least two occupied neighbors become occupied. While bootstrap percolation is typically defined in discrete time, the final configuration of the network is independent of whether the dynamics happens in 
discrete or continuous (as in our model) time. 

We will use random graphs as models for complex networks on which the QCP is taking place. We will denote by $d$ the degree of a randomly chosen vertex in the network and the degree distribution by $p_k = \mathbb{P}(d=k)$.
We are interested in networks with size $n \rightarrow \infty$ and where the vertex degrees are uncorrelated. The specific random graphs that we will consider are 
\begin{itemize}
 \item \emph{Random regular graphs} RR($\mu$) in which each vertex has degree $\mu$. Since everyone has exactly $\mu$ friends, this graph is not a good model of a social network. However, the fact that it looks locally like a tree will facilitate proving results.
\item \emph{Erd\H{o}s-R\'{e}nyi random graphs} ER$(\mu)$ where each pair of vertices is connected with probability $\mu/n$. In the $n\rightarrow \infty $ limit, the degree distribution of the limiting graph is Poisson with mean $\mu$. This is a 
prototypical model for the situation in which the degree distribution has a rapidly decaying tail.
\item \emph{Power law random graphs} PL($\alpha$) with degree distribution $ p_k = c k^{-\alpha}$. We are particularly interested in graphs where the exponent $\alpha$ lies between 2 and 3, which has been found to be the case for many real world networks \cite{barabasi03}. 
\end{itemize}
We will occasionally refer to RR and ER as homogeneous networks as their degree distributions are peaked around the mean, in contrast to PL where the distribution has a heavy tail.

\section{Mean field calculations} \label{mean}
We can attempt an analytical study of the dynamics by writing the equations for the various \emph{moments} of the network. Let $g$ be a small graph 
labeled with 1's and 0's. We define the $g-$moment, written as $\langle g\rangle $, of a $\{0,1\}$ valued process on a graph $G$ as the expected number of copies of $g$ that 
exist in the set of all subgraphs of $G$. For example if $g = 1-0-1 $, we look at all the connected vertex triples in the network and count the ones 
where the center vertex is in state 0 and the other two vertices are in state 1. We will write $ \rho(\lambda,\rho^{(0)};t) $  as the density 
$\langle 1
\rangle /n$ at time $t$ with a birth rate of $\lambda$ and an initial configuration where each vertex is independently occupied with a 
probability $\rho^{(0)}$. The order parameter for our phase transitions is the steady state density 
\begin{equation}
 \rho_*(\lambda,\rho^{(0)}) = \lim_{t\rightarrow \infty} \rho(\lambda,\rho^{(0)};t)\,.
\end{equation}
We define the critical birth rate $\lambda_c$  as the birthrate above which there exists a stable steady state density that is greater than zero, i.e., 
\begin{equation} 
 \lambda_c = \inf\{ \lambda :    \rho_*(\lambda,1) >0  \}\,.
\end{equation}
In the definition above, we chose $\rho^{(0)}=1$ since it has the best chance of having a positive limit. 
We also define a critical initial density $\rho_c$ 
as the minimum initial density required to reach a positive steady state density
when the birth rate is infinite, i.e., 
\begin{equation}
  \rho_c = \inf\{ \rho^{(0)}:  \lim_{\lambda \rightarrow \infty}
\rho_*(\lambda,\rho^{(0)}) > 0 \}\,. \label{kirk}
\end{equation}

From their definitions, it is straight forward to write the dynamical equations
of $\langle 1 \rangle$ for the VQCP and the EQCP, 
\begin{equation}
\frac{\D }{\D t} \langle 1 \rangle = - \langle 1 \rangle + \lambda  
\begin{cases}
                                             \langle 1 - 0 - 1 \rangle &
\textrm{ for the VQCP}\\
					    \langle 1 - 1 - 0 \rangle & \textrm{
for the EQCP}\\
                                            \end{cases} \,.
 \label{firstmoment}
\end{equation}

If we were to write the equations for the third order moments that appear on the
RHS of \eqref{firstmoment}, those equation would involve still higher order
moments. 
Continuing this way, we end up with an infinite series of equations that are not
closed. Therefore we resort to a mean field approximation by assuming the states of neighbors of a vertex to be independent at all times.  

\subsection{Homogeneous networks} \label{short}
\begin{figure}
\centering
\includegraphics[width=8.6cm]{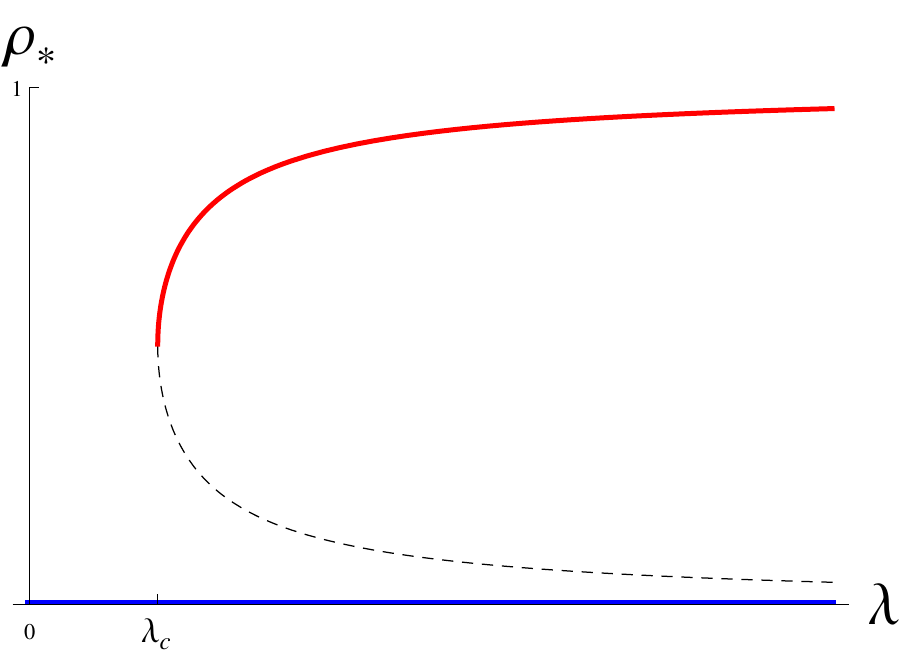}  
\caption{The blue, dashed and red curves correspond to $\rho_{*} = 0$, $\rho_-$
and $\rho_+$ respectively obtained from the mean field calculation for both QCP types on homogeneous networks.}\label{SimpleMeanField}
\end{figure}

In the following we do a naive calculation that ignores the correlation between degree and occupancy, which should be reasonable for homogeneous networks. 
With these assumptions, $\langle 1-0-1\rangle$ will be $ n \rho^2 (1-\rho)
\langle \binom{d}{2}\rangle$. Plugging this value into \eqref{firstmoment} we
get
\begin{equation}
\dot{\rho} = - \rho + \lambda  \rho^2 (1-\rho) \left\langle \binom{d}{2} \right
\rangle\,.
\label{SimpleMF}
\end{equation}
Setting the RHS of \eqref{SimpleMF} to zero gives a cubic equation whose roots
are
the
possible steady state densities $\rho_{*}$. 
Clearly, zero is a trivial root of \eqref{SimpleMF}. The other two
roots are
\begin{equation}
\rho_{\pm} = \frac{1}{2} \left[  1    \pm \sqrt{1 - \frac{\lambda_c} { \lambda 
}} \right] \, .
\end{equation}
These solutions are real only 
when $\lambda > \lambda_c  = 4/\langle \binom{d}{2}  \rangle$. 
In the language of nonlinear dynamics, \eqref{SimpleMF} exhibits a saddle node
bifurcation at $\lambda_c $. It is easy to see that zero and $\rho_+$ are stable
fixed points whereas $\rho_-$ is an unstable fixed
point. This can be seen in \fig{SimpleMeanField}{}. 
The limiting critical initial density is 
\begin{equation} \label{mandarin}
 \rho_c = \lim_{\lambda\rightarrow \infty} \rho_- = 0 \,.
\end{equation}

For ER($\mu$) we have $\langle \binom{d}{2} \rangle  = \mu^2/2$ which gives $\lambda_c  = 8/\mu^2$. For PL($\alpha \leq 3$) we have $\langle \binom{d}{2} \rangle = \infty$ so $\lambda_c =0$ while 
PL($\alpha>3$) has finite $\langle \binom{d}{2} \rangle$ leading to a non-zero value for $\lambda_c $.  
The mean field calculation for the EQCP is essentially the same as done above
and predicts the same qualitative features. Thus, for networks with finite $\langle \binom{d}{2} 
\rangle$, the simple mean field calculation predicts a discontinuous phase transition at $\lambda =\lambda_c $
and a region of bistability for $\lambda > \lambda_c $, for both
QCP types.

\subsection{Heavy tailed degree distributions}
The mean field calculation of Section \ref{short} is simplistic since it ignores the fact that the occupancy probability depends on the degree.
Pastor-Satorras and Vespignani \cite{pastor01} improved the mean field approach for the
linear contact process 
by defining $\rho_k$, the fraction of vertices of degree $k$ that are occupied, and $\theta$, the probability that a given edge points to 
an occupied vertex. These variables can be related through the size biased degree distribution $q_k =  k p_k / \langle d \rangle$ which is the distribution of the degree of a vertex 
at the end of a randomly chosen edge. 
\begin{equation}
\theta
= \sum_k q_k \rho_k \label{davos} 
\end{equation}
Note that for homogeneous networks we assumed $\theta
=\rho$. 
As before, the state of the neighbors of a vacant vertex are assumed to be
independent. So the number of occupied neighbors of a vertex of degree $k$
follow 
a distribution $\mathrm{Binomial}(k,\theta)$. This enables us to apply this approach to the VQCP. We write equations for $\rho_k$,
\begin{equation}\dot{\rho}_k = -\rho_k + \lambda (1-\rho_k)    \binom{k}{2}
\theta^2 \, .\end{equation} So in steady state 
\begin{equation} \label{geneva}
\rho_{k*} =  \frac{\lambda \binom{k}{2} \theta_*^{2}}{1+ \lambda \binom{k}{2}
\theta_*^{2}}\,.
\end{equation}
Combining \eqref{davos} and \eqref{geneva} leads us to a self-consistent equation for $\theta_*$.

\begin{equation}
\theta_{*} = \theta_{*} I(\lambda, \theta_{*}) \, , \label{asia}
\end{equation}
where 
\begin{equation}
 I(\lambda,\theta) = \sum_{k}^\infty \frac{k p_k}{\langle d \rangle} \left[ 
\frac{\lambda \binom{k}{2} \theta}{1+ \lambda \binom{k}{2} \theta^2}\right] \,.
\label{dubai}
\end{equation} 
Clearly, $\theta_{*} = 0$ is a solution of \eqref{asia}. Finding a non-trivial
solution involves solving 
\begin{equation}
I(\lambda, \theta_{*})=1 \quad ,\quad  \theta_*\in (0,1) \,. \label{europe}
\end{equation}
\begin{figure}
\centering
\includegraphics[width=8.6cm]{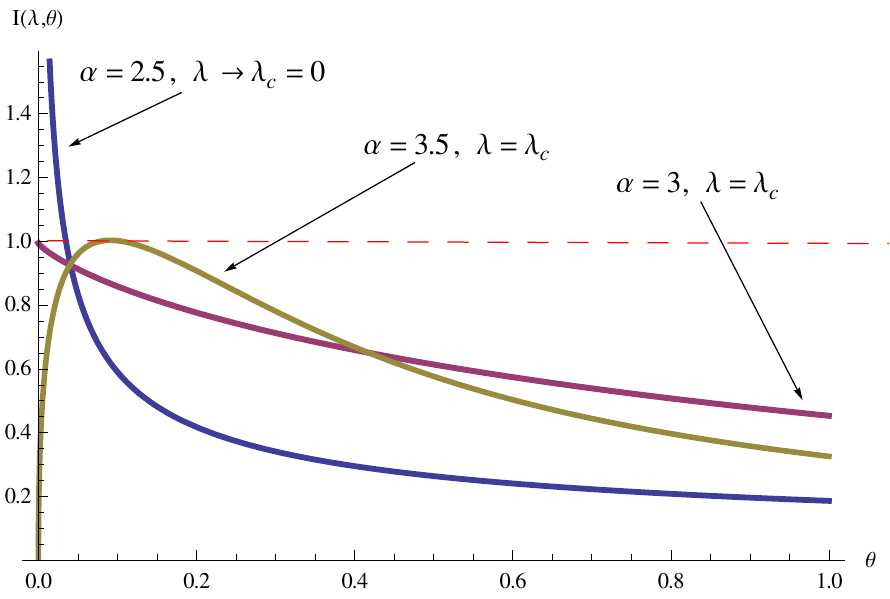}  
\caption{$I(\lambda,\theta)$ versus $\theta$ near $\lambda=\lambda_c$ for various power law graphs.}\label{I_theta}
\end{figure}
For power law graphs PL($\alpha$), as we will now show, the mean field calculation predicts
\begin{itemize}
  \item If $\alpha > 3$, $\lambda_c >0$ and the transition is discontinuous.
  \item If $\alpha = 3$, $\lambda_c >0$ and the transition is continuous.
  \item If $2< \alpha < 3$, $\lambda_c=0$ and the transition is continuous.
\end{itemize}   
To begin we note that if $p_k \sim Ck^{-\alpha}$ with $\alpha>4$ then 
\begin{equation}
I(\lambda,\theta) \le \frac{\theta \lambda}{2\langle d \rangle} \sum_k Ck^{3-\alpha} \to 0 
\end{equation}
as $\theta \to 0$. When $3< \alpha \le 4$ breaking the sum at $k=1/\theta$, and bounding the fraction
by $C/\theta$ for $k\geq 1/\theta$ (here and in what follows $C$ is some finite constant).
\begin{equation}
 I(\lambda,\theta) \le  \frac{\theta \lambda}{2\langle d \rangle} \sum_{k=1}^{1/\theta} Ck^{3-\alpha}
+ \frac{1}{\theta \langle d \rangle} \sum_{k=1/\theta}^\infty C k^{1-\alpha} \to 0
\end{equation}
as $\theta \to 0$. Since $I(\lambda,1)<1$ for all $\lambda$, the curve will have $\sup_{\theta\in[0,1]}I(\lambda_c,\theta) =1$
at some $\lambda_c$. Then for $\lambda>\lambda_c$ there will be two roots, the larger of which
is the relevant solution.

When $2< \alpha \le 3$ approximating the sum by an integral we have
\begin{equation}
I(\lambda,\theta) \sim \theta^{\alpha-3} \int_0^\infty x^{1-\alpha} \frac{\lambda x^2}{1+\lambda x^2} \, \D x \,.
\end{equation}
From this we see that if $2<\alpha < 3$ then as $\theta \to 0$, $I(\lambda,\theta) \to \infty$, so there is a solution 
to $I(\lambda,\theta_\lambda)$ for any $\lambda>0$, and $\theta_\lambda \to 0$
as $\lambda\to 0$. For
small values of $\rho_{*}$, a little calculation shows that the following relation holds
\begin{equation}
\rho_{*} \sim (\lambda - \lambda_c)^{\gamma(\alpha)} \quad \textrm{ where }
\gamma(\alpha) =  \frac{1}{3-\alpha} -\frac{1}{2} \, \label{africa}
\end{equation}
is the critical exponent.

In the borderline case $\alpha=3$ the limit as $\theta \to 0$ is finite. 
There is a critical value $\lambda_c$ so that $I(\lambda_c,0)=1$. Since $\theta \to I(\lambda,\theta)$
is decreasing, it follows that 
for $\lambda > \lambda_c$ we have one solution $I(\lambda,\theta_\lambda)=1$ 
which has $\theta_\lambda \to 0$ as $\lambda\to\lambda_c$. The behavior of $I(\lambda\approx \lambda_c,\theta)$ is shown in \fig{I_theta}.

A second way to determine the nature of the phase transition is to adapt the argument
of Gleeson and Cahalane \cite{gleeson07}, which can be applied if we use a discrete time version of
the model in which a vertex with $k$ neighboring pairs will be occupied at the next step with probability
$1 - (1-p)^k$. The computation in their formulas (1)--(3) supposes that the vertices at a
distance $n$ from $x$ are independently occupied with probability $\rho_0$. The function
$G(\rho)$ defined in their (3) gives the occupancy probabilities at distance $k-1$ assuming that
the probabilities at distance $k$ are $\rho$.  Iterating $G$ $n$ times and letting $n\to\infty$
gives a prediction about the limiting density in the cascade. if one repeats the calculation for
our system then 0 is an unstable fixed point when $\alpha < 3$, while it is locally attracting for
$\alpha > 3$. This agrees with the mean-field prediction of $\lambda_c=0$ in the former case
and a discontinuous transition with $\lambda_c>0$ in the second.

\section{Some Rigorous results} \label{rigor}
We have not been able to extend the mean field calculation to the EQCP on power law graphs, but by generalizing an argument of Chatterjee and Durrett \cite{chatterjee09} we can 
prove that $\lambda_c=0$ for $\alpha \in (2,\infty)$. The details are somewhat lengthy, so we only explain the main idea. Consider a tree in which the vertex 0 has $k$ neighbors and each of 
its neighbors has $l$ neighbors and $l$ is chosen so that $l\lambda \geq 10$. One can show that if $k$ is large then with high probability the infection will persist on this graph for time $\geq \exp(c(\lambda) k)$. 
In a power law graph one can find such trees with $k=n^{1/(\alpha-1)}$. Using the prolonged persistence on these trees as a building block one can easily show that if we start with all vertices occupied the infection 
persists for time $\geq \exp(n^{1-\epsilon})$ with a positive fraction of the vertices occupied. With more work (see \cite{mountford11,mountford12}) on can prove persistence for time $\exp(c(\lambda) n)$. 

For both types of QCP it is easy to show that it is impossible to have a
discontinuous transition with $\lambda_c =0$. 
The proof for VQCP is as follows. Let $\langle 1_k\rangle$ be the expected number of occupied
sites of degree $k$ and $\langle 10_k 1 \rangle$ be the expected number of 1-0-1 triples
when the 0 vertex has degree $k$. We can write an equation similar to \eqref{firstmoment}
\begin{equation}
\frac{\D }{\D t} \langle 1_k\rangle = - \langle 1_k\rangle + \lambda  \langle1 -
0_k - 1\rangle
\end{equation}
which means at steady state 
\begin{eqnarray}
\langle 1_k\rangle_* = \lambda  \langle1 - 0_k - 1\rangle_* \leq \lambda 
\langle0_k\rangle_* \binom{k}{2} \nonumber\\ 
\Rightarrow \rho_{k*} \leq \lambda (1-\rho_{k*})
\binom{k}{2} \Rightarrow \rho_{k*} \leq \frac{\lambda \binom{k}{2}}{1+\lambda
\binom{k}{2}} \,.
\end{eqnarray}
So, as $\lambda \rightarrow 0$, $ \rho_{k*} \rightarrow 0$ and $\rho_* = \sum_k
\rho_{k*} p_k \rightarrow 0$. Thus the transition will be continuous.
The proof for EQCP is similar. In that case the subscript $k$ stands for the
secondary degree $d^{(2)}$ which is defined as the number of neighbors of
neighbors of a given vertex (not including itself), i.e., 
$d^{(2)}(x) = |\{z : z \sim y, y\sim x, z \neq x\}|$.  
\begin{equation}
\frac{\D }{\D t} \langle 1_k\rangle = - \langle1_k\rangle + \lambda  \langle1 -
1-0_k\rangle
\end{equation}
So at steady state 
\begin{equation}
\langle1_k\rangle_* = \lambda  \langle1 - 1-0_k\rangle_* \leq \lambda \langle0_k\rangle_* k 
\Rightarrow \rho_{k*} \leq \frac{\lambda k}{1+\lambda k} 
\end{equation}
Thus for both QCP types we find that if $\lambda_c=0$ then the phase transition is continuous.

For both QCP types on random $r$-regular graphs we can show that the critical birth rate is positive as follows. In the EQCP let there be $m$ occupied vertices. Each of these $m$ vertices can have at most $r$ neighbors that 
are vacant and can give birth on to them at rate $\leq (r-1)\lambda $ or die at rate 1. So the total birth rate in the network is $\leq (r-1)\lambda  r  m$ against a death rate of $m$,
 and it follows that $\lambda_c > 1/r(r-1)$. Similarly, for the VQCP the total birth rate is 
$\leq \lambda \binom{r}{2} r m$ and it follows that $\lambda_c > 1/ r\binom{r}{2} $. These arguments depend on the degree being bounded, so they do not work for Erd\H{o}s-R\'{e}nyi and power law graphs. 

\section{Simulation results} \label{simulation}
We perform simulations of the QCP on RR(4), ER(4) and PL(2.5). We generate the random regular and power law random graphs using the recipe called \emph{configuration model} \cite{molloy95}. 
We draw samples $d_x$ from the degree distribution and attach that many ``half-edges'' to vertex $x$. We 
pair all the half edges in the network at random. We then delete all self loops and multiple edges. When $\alpha>2$ this does not significantly modify the degree distribution.
If $\sum_x d_x $ turns out to be odd (an event with probability $\approx \frac{1}{2}$), we ignore the last remaining unpaired half-edge. Furthermore, for PL(2.5) we start the degree distribution at 3 as, in the VQCP, the vertices of degree 1 and 2 are impossible 
or difficult to get occupied.

\begin{figure}
\includegraphics[trim=1.5cm 1cm 1cm 1cm, width=9.5cm]{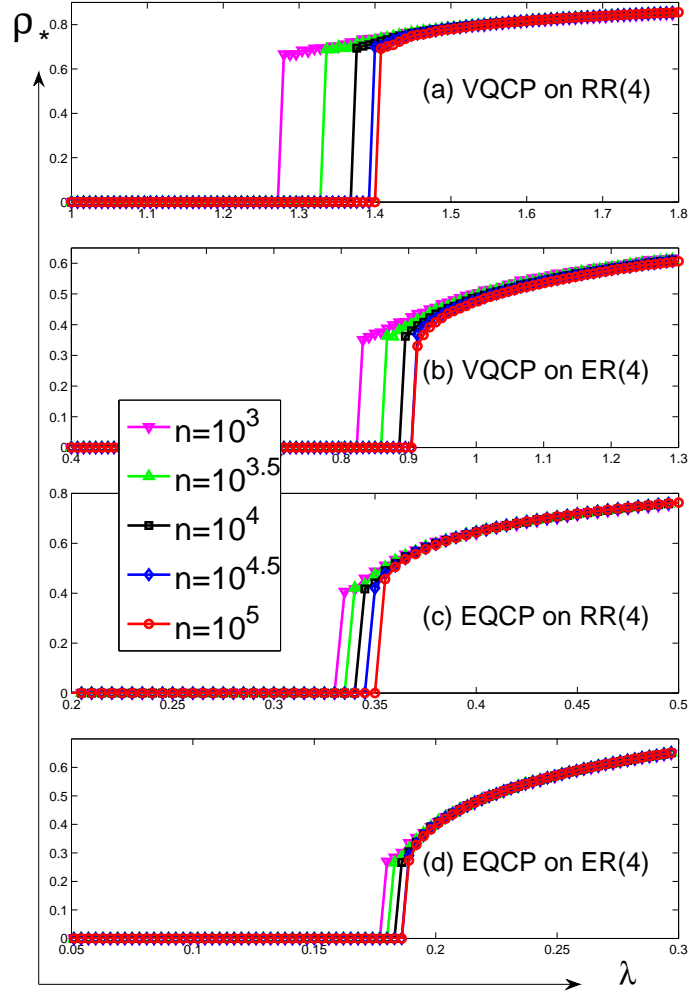}  
\caption{Steady state density reached, starting from all vertices occupied, for QCP on homogeneous networks of various sizes $n$.\label{VRr1VEr1ERr1EEr1}}
\end{figure}

\begin{figure}
\includegraphics[trim=1.5cm 2cm 1cm 0cm,width=9.3cm]{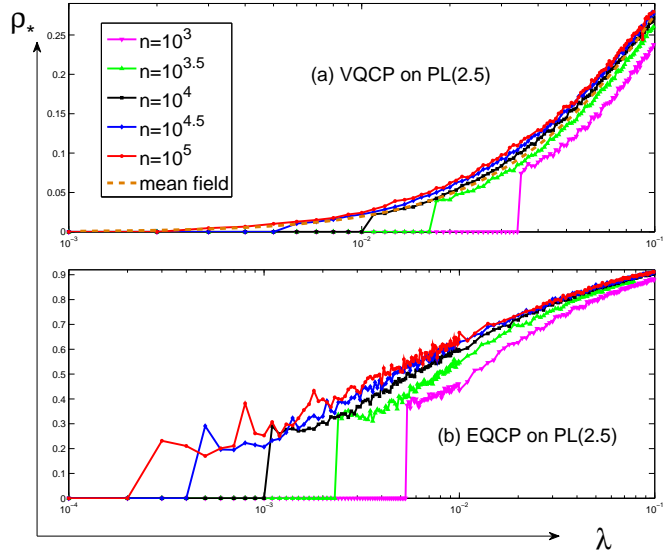}  
\caption{Steady state density reached, starting from all vertices occupied, for QCP on power law networks of various sizes $n$.
Note that the $\lambda$ axis is in the log scale.}\label{VPr1EPr1}
\end{figure}

To deal with finite size effects, we observe how the plot of the steady state density $\rho_{*}(\lambda,1)$ versus $\lambda$ starting with all vertices occupied changes 
when size $n$ of the network ranging from $10^{3}$ to $10^5$. \fig{VRr1VEr1ERr1EEr1}{} shows the results of both QCP types on RR(4) and ER(4). Here 
the curves seem to converge to a positive value implying a positive $\lambda_c$. 
The results for PL(2.5) are shown in \fig{VPr1EPr1}{}. We observe that the transition happens close to zero and moves towards zero with increasing $n$ indicating that the 
critical birth rate is zero. As explained earlier, if $\lambda_c=0$ then the transition is continuous. This is consistent with the the mean field predictions for the VQCP and rigorous result for EQCP. In addition, in \fig{VPr1EPr1}{(a)}, the critical exponent for the $n=10^5$ 
curve can be measured to be approximately 1.45 which is close to the mean field value of 1.5 (obtained by setting $\alpha =2.5$ in \eqref{africa}).

\begin{figure}
\includegraphics[trim=2cm 1cm 0cm 1cm,width=9.5cm]{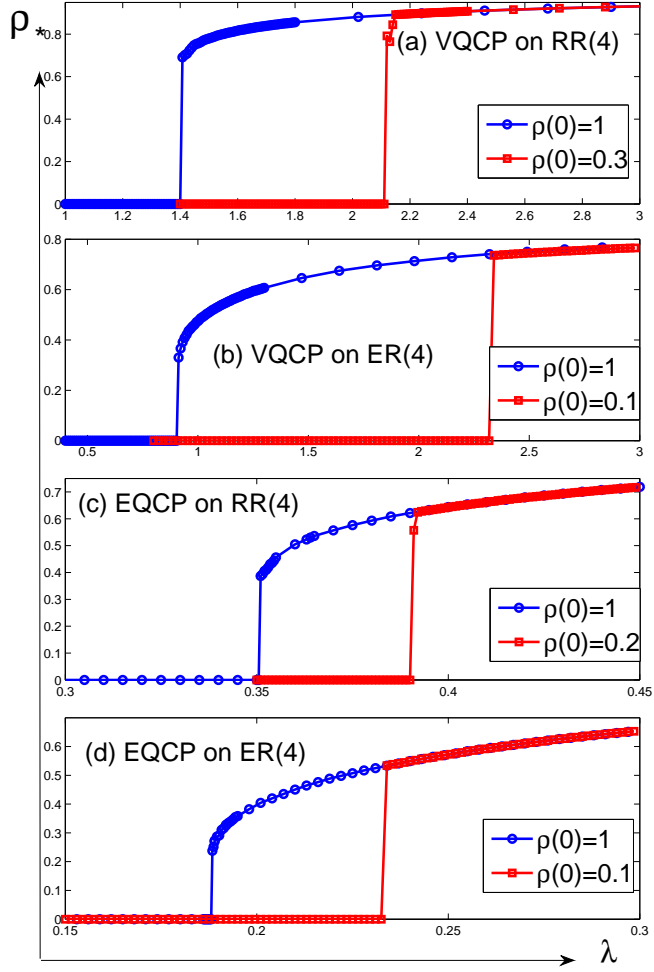}  
\caption{Steady state density reached, starting from two different initial densities $\rho^{(0)}$, for QCP on homogeneous networks of size $n=10^5$. Notice the similarity with the mean field prediction on \fig{SimpleMeanField}.\label{VRn5VEn5ERn5EEn5}}
\end{figure} 

In order to further investigate the phase transitions in random regular and Erd\H{o}s-R\'{e}nyi graphs, we look at the steady state density attained by starting from 
two different initial densities for the same network size $n=10^5$. \fig{VRn5VEn5ERn5EEn5}{} again shows a similar pattern across both QCP and 
both network types. In \fig{VRn5VEn5ERn5EEn5}{(b)} we see that for birth rates between 0.9 and 2.3, the VQCP survives when the starting configuration had all vertices 
occupied but dies out when starting with only one-tenth of the vertices occupied. Thus we see bistability in the region $\lambda \in (0.9,2.3)$ implying 
a discontinuous transition and consequently that $\lambda_c$ is positive and close to 0.9. This is qualitatively in agreement with the mean field prediction seen in \fig{SimpleMeanField}, although the critical birth 
rate of 0.9 shows a deviation from the mean field value of $8/4^2=0.5$.

\begin{figure}
\includegraphics[trim=2.4cm 2cm 1cm 2cm,width=9.5cm]{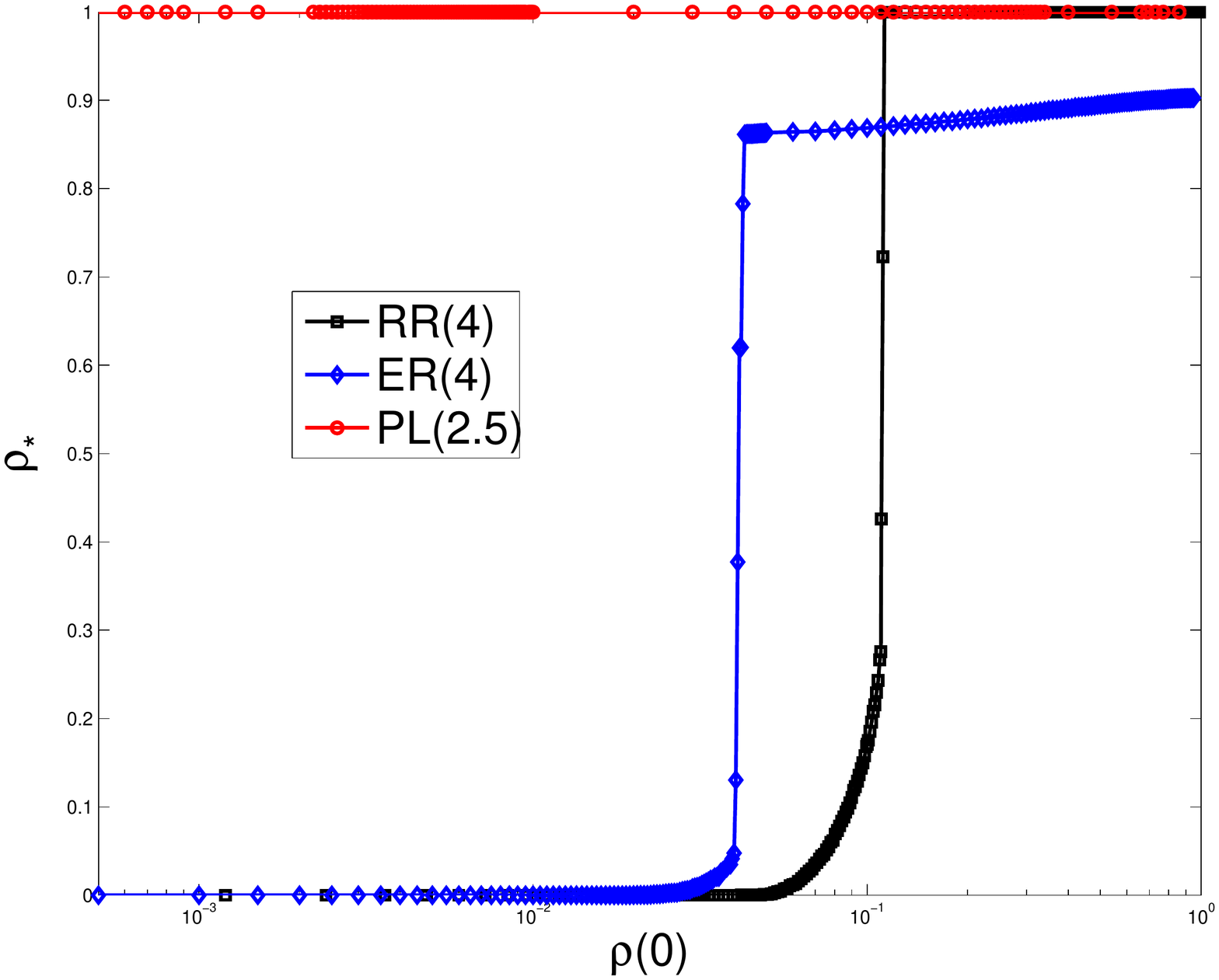}  
\caption{Steady state density when the birth rate is infinite in the VQCP. Note that the $\rho^{(0)}$ axis
is in the log scale.}\label{Vn5bInf}
\end{figure}

Fontes and Schonmann \cite{fontes08} have shown that for bootstrap percolation on the tree there is a critical density
$p_c$ so that if the initial density is $< p_c$ then the final bootstrap percolation configuration 
has no giant component of occupied sites. In this situation having deaths at a positive
rate in the VQCP will lead to an empty configuration. The last argument is for the tree, but results of Balogh and Pittel \cite{balogh07}
show that similar conclusions hold on the random regular graph. While this argument is not completely rigorous,
the reader should note that since all of the VQCP are dominated by bootstrap percolation, it follows that the limiting critical initial density defined in \eqref{kirk} has $\rho_c > 0$ 
in contrast to the mean field prediction in \eqref{mandarin}. \fig{Vn5bInf}{} shows the final density attained as a function of the initial density when the birth rate is infinite (and death rate is positive). 
We see that $\rho_c $ in the VQCP is positive for the random regular and Erd\H{o}s-R\'{e}nyi graphs whereas it is zero for the power law graph. 
The corresponding results (not shown here) in the case of the EQCP indicate that $\rho_c =0$ for random regular, Erd\H{o}s-R\'{e}nyi and power law random graphs.

\section{Conclusion}
\label{conclusion}
\begin{table}
\centering
\begin{tabular}{|l||l|l|l|}
\hline
  & Linear CP & Vertex QCP & Edge QCP \\  \hline \hline
 1D  & cont.\, , $+$ \cite{durrett80,liggett99}& NA \, , $\infty$ & cont.\, , $+$ \cite{prakash97}\\  \hline
2D  & cont.\, , $+$ \cite{bezuidenhout90,liggett99}& discon.\, , $+$ \cite{liu09}& cont.\, , $+$ \cite{grassberger82}\\  \hline  
RR & cont. \cite{morrow94}\, , $+$ \cite{pemantle92} & discon.\, , $+$ & discon.\, , $+$ \\  \hline  
ER & cont.\, , $+$ \cite{parshani10}& discon.\, , $+$ & discon.\, , $+$ \\  \hline  
PL((2,3)) & cont.\, , $0$  \cite{chatterjee09}& cont.\, , $0$ & cont.\, , $0$ \\  \hline  
PL(3) & cont.\, , $0$ \cite{chatterjee09}& cont.\, , $ +$ & cont.\, , $0$ \\  \hline
PL((3,$\infty$)) & cont.\, , $0$ \cite{chatterjee09}& discon.\, , $+$ & cont.\, , $0$ \\  \hline    
\end{tabular}
\caption{Nature of phase transitions of contact processes on various networks. Note that `$0$',`$+$' and `$\infty$' stand for zero, positive and infinite values respectively of $\lambda
_c$.}
\label{PTtable}
\end{table}
In this paper we have investigated the properties of two versions of the quadratic contact process on three types of random graphs. The mean field calculations we performed agree with the simulation results. This may be due to the fact that complex networks have exponential volume growth and therefore are like infinite dimensional lattices where mean field is exact.

Table \ref{PTtable} summarizes what is known about the phase transitions of contact processes in 1 and 2 dimensional lattices and on the random graphs RR, ER and PL. The positivity of the critical birth rate for 1D, 2D and RR follows trivially from the boundedness of their degrees. For VQCP on a 1D lattice, two consecutive 0's can never get filled and it follows that $\lambda_c = \infty$. The results for the linear process on RR are inferred from the rigorous results for trees and the fact that RR is locally tree like.

The results indicate that the EQCP is qualitatively not very different from the linear contact process on low dimensional lattices and power law graphs, in contrast to the VQCP, which differs from its low dimensional analogue. In view of the fact that they are very different in how they fill vacant vertices on a network, the similarity between VQCP and EQCP in their phase transitions on complex networks is a little perplexing. 

The EQCP can easily propagate on a chain and ``cross bridges'' connecting communities, compared to the VQCP which always requires two occupied neighbors. In the EQCP vertices with a large number of neighbors of large degree are the key to its survival. However, in the VQCP it is impossible for the central vertices to repopulate the leaves, so these structures are not long lasting. In contrast the Gleeson-Cahalane calculation suggests that survival is due to the fact that as waves of particles move through the system the densities increase.

It is also interesting to note that the discontinuous phase transitions in the QCP are accompanied by regions of bistability. Such a feature seems to be typical of all non-equilibrium phase transitions \cite{muller97,monetti01}. 

\bibliography{myrefs}

\end{document}